%% file: paper.tex
\documentclass{IEEEtran}

\usepackage{caption}

\usepackage{cite}

\ifCLASSINFOpdf
\else
\fi

\usepackage{amsmath}

\usepackage{url}

\usepackage{amssymb}
\usepackage{xspace}
\usepackage{enumitem}
\usepackage{pgfplots}
\usetikzlibrary{shapes.geometric}

\newcommand{\ie}{\emph{i.e.,}\xspace}
\newcommand{\eg}{\emph{e.g.,}\xspace}

\newcommand{\viz}{\emph{viz.,}\xspace}
\newcommand{\etal}{\emph{et al.}\xspace}

\newcommand{\Aut}{\operatorname{Aut}}

\usepackage{amsthm}
 
\theoremstyle{plain}
\newtheorem{theorem}{Theorem}
\newtheorem{definition}{Definition}
\newtheorem{maintheorem}[theorem]{Main Theorem}
\newtheorem{lemma}{Lemma}

\newtheorem{remark}{Remark}
\newtheorem{example}{Example}

\newcommand{\textdiff}[1]{\ensuremath{\Delta_{\mbox{\normalfont \footnotesize {#1}}}}}
\newcommand{\tdiff}[0]{\textdiff{T}}
\newcommand{\sdiff}[0]{\textdiff{S}}

\renewcommand{\emptyset}{\varnothing}

\newcommand{\dhl}[1]{}

\begin{document}
\title{A Graph Symmetrisation Bound on  Channel Information Leakage under Blowfish Privacy}

\author{Tobias~Edwards,
        Benjamin~I.~P.~Rubinstein,
        Zuhe~Zhang,
        and~Sanming~Zhou%
\thanks{Manuscript received July 10, 2020; accepted September 29, 2021. This work was supported in part by the ARC through Grant DE160100584.}%
\thanks{T. Edwards and B. Rubinstein are with the School of Computing and Information Systems, University of Melbourne, Parkville, VIC 3010, Australia (email: tobias.w.edwards@gmail.com, brubinstein@unimelb.edu.au).}%
\thanks{Z. Zhang is with the Australian Tax Office, Docklands, VIC 3008, Australia (email: zhang.zuhe@gmail.com).}%
\thanks{S. Zhou is with the School of Mathematics and Statistics, University of Melbourne, Parkville, VIC 3010, Australia (email: sanming@unimelb.edu.au).}}

\markboth{IEEE TRANSACTIONS ON INFORMATION THEORY,~Vol.~X, No.~X, X~2021}%
{Edwards \MakeLowercase{\textit{et al.}}: Blowfish Privacy Implies Bounded Min-Entropy Leakage}

\maketitle

\begin{abstract}
Blowfish privacy is a recent generalisation of differential privacy that enables improved utility while maintaining privacy policies with semantic guarantees, a factor that has driven the popularity of differential privacy in computer science. This paper relates Blowfish privacy to an important measure of privacy loss of information channels from the communications theory community: min-entropy leakage. Symmetry in an input data neighbouring relation is central to known connections between differential privacy and min-entropy leakage. But while differential privacy exhibits strong symmetry, Blowfish neighbouring relations correspond to arbitrary simple graphs owing to the framework's flexible privacy policies. To bound the min-entropy leakage of Blowfish-private mechanisms we organise our analysis over symmetrical partitions corresponding to orbits of graph automorphism groups. A construction meeting our bound with asymptotic equality demonstrates tightness.
\end{abstract}


\IEEEpeerreviewmaketitle

\section{Introduction}
\IEEEPARstart{D}{ifferential} 
privacy~\cite{Dwork1} has emerged as a leading measure of privacy 
loss across the machine learning, theoretical computer science, databases and computer security communities. 
Its success is due in large part to strong guarantees on the indistinguishability of
input datasets based on releases of randomised mechanisms such as
learned models~\cite{abadi2016deep,bassily2014private} and derived data structures~\cite{cormode2012differentially} over sensitive data.
This indistinguishability takes the form of mechanism response
distribution smoothness over pairs of adjacent
datasets---those that differ in one record.
By relaxing the differential privacy adjacency relation, 
randomised mechanisms may achieve higher utility for the same
privacy level on select pairs of datasets. The Blowfish 
framework, introduced by He \etal~\cite{Blowfish}, attains this goal while
maintaining meaningful privacy policies. As a result, Blowfish
adjacency relations may lack the symmetry
of those under differential privacy.

Our goal in this paper is to examine Blowfish privacy and 
its relationship to min-entropy leakage~\cite{QuantInfoFlow}---a leading notion of privacy in communications theory.
Specifically we establish that bounded Blowfish privacy implies
bounded min-entropy leakage. Together with previous
work~\cite{Blowfish, MSAlvim1, MSAlvim3, MSAlvim2} this completes
the following strict hierarchy 
(with the converse implications not holding~\cite{Blowfish,MSAlvim2}).

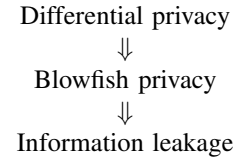
\begin{center}
    \text{Differential privacy} \\
    $\Downarrow$ \\
    \text{Blowfish privacy} \\
    $\Downarrow$ \\
    \text{Information leakage}
\end{center}

The dataset adjacency relation can be viewed as a simple graph.
Previous work bounding information leakage of differentially-private
mechanisms exploit strong symmetry assumptions of this
graph. %
While this corresponds to distance regularity and vertex transitivity, 
Blowfish-induced graphs can be arbitrary. 
As a result the challenge for analysing Blowfish privacy is one of graph symmetrisation. 
Our main bounds accomplish this by developing a new proof technique that organises the
graph by vertex-transitive automorphism orbits. Beyond demonstrating a meaningful 
connection, we discuss implications of our results on understanding of both Blowfish
and information-leakage frameworks.

Before describing our main results, we overview related work and describe necessary background material in differential privacy (Section~\ref{sec:bg}). Our presentation of Blowfish privacy is greatly simplified over the original exposition~\cite{Blowfish}, and thus may be of independent interest. 

\subsection{Related Work}

While differential privacy's success is owed in large part to its worst-case
guarantees, researchers have sought natural relaxations that: improve utility
while maintaining semantic privacy guarantees, offering generic mechanisms, and permitting mechanism composition. 

Approximate $(\epsilon,\delta)$-differential privacy---the most well-known
variant---relaxes pure $\epsilon$-differential privacy response distribution 
smoothness, on low-probability responses~\cite{Dwork1}. In so doing, it
permits guarantees on privacy loss for highly concentrated mechanisms such as the
Gaussian~\cite{nikolov2013geometry}. Citing the ensuing unbounded residual
privacy risk on tails, %
Mironov~\cite{mironov2017renyi} proposed R\'enyi differential privacy, 
based on R\'enyi divergence, to generalise (approximate) differential
privacy while bounding tails of the privacy loss random variable. An alternate approach
based on bounding all moments of the privacy loss variable, improving 
rates for composition of approximate DP, is concentrated differential
privacy (CDP)~\cite{dwork2016concentrated}.

Noting that pathological datasets can contribute to high query sensitivity, and so 
high utility loss, Hall \etal~\cite{hall2012random} introduced random differential privacy
which requires response distribution smoothness to hold not
on all datasets but rather on i.i.d. datasets with high probability. Their framework permits analysis 
of mechanisms run on unbounded input data, for example, and 
permits private release under estimation of sensitivity of black-box
functions~\cite{rubinstein2017pain}.

Kifer and Machanavajjhala~\cite{Pufferfish} proposed the Pufferfish privacy framework to provide
privacy guarantees in the face of varying threat models. Notably the framework
accounts for prior releases
of non-differentially private information. Kifer and Machanavajjhala~\cite{NoFreeLunch} demonstrate that without such
an extension, large amounts of sensitive information may be leaked.

Inspired by Pufferfish, and a focus of this paper, is the Blowfish framework
introduced by He \etal~\cite{Blowfish}. As detailed in Section~\ref{sec:bg-blowfish}, the
approach taken is for the defender to define a subset of data values to keep secret,
as well as constraints on data already known publicly. These secrets and constraints 
together induce the adjacency relation on which response distribution smoothness is 
(relaxed) to hold on. In this way the generalised Blowfish threat model is parametrised by a semantic privacy policy. While Blowfish privacy adopts the smoothness criterion on
response distributions of differential privacy, relaxations including R\'{e}nyi DP and CDP
are built on the same adjacency relationship.

The communications theory community have also developed frameworks for
guaranteeing privacy, exemplified by the study of quantitative information flow~\cite{QuantInfoFlow} which characterises how information channels leak information
with change to distributional entropy (\viz Section~\ref{sec:infoflow}).
We continue to study the connections between quantitative information flow and differential privacy as initiated by Alvim \etal~\cite{MSAlvim1}. 
Other researchers have followed this thread of work also. 
For example Dwork \etal highlight early connections between differential privacy and relative entropy~\cite{dwork2010boosting}.
More recently, Issa \etal~\cite{issa2019operational} situate local differential privacy~\cite{duchi2013local} within a guessing framework designed for interpreting leakage definitions. 
In a celebrated result of practical significance, reformulating differential privacy as max-divergence admits an application of Azuma's inequality to bound differential privacy of adaptive compositions of mechanisms~\cite{DworkBook}. 
Independent of Alvim \etal, Barthe and K\"{o}pf developed bounds on the leakage of differentially-private mechanisms, without using the same symmetry properties leveraged by Alvim \etal. For mechanisms acting on binary $n$-strings, they achieve a bound of $n \epsilon \log_2(e)$~\cite[Corollary~2]{barthe2011information}, which we recover in this paper. They then go on to improve their bound by exploiting specific structure of differential privacy that does not hold for Blowfish privacy in general. Our setting applies to more general input data, and makes fewer assumptions about the database neighbouring relation owing to the flexibility of Blowfish.

\section{Background}\label{sec:bg}

We next recall the Blowfish and information flow frameworks.

\subsection{Blowfish Privacy}\label{sec:bg-blowfish}

Adopting the language of differential privacy~\cite{Dwork1} from statistical databases, we consider
a {\it database} $D$ as comprising $n$ {\it records} each taking a value in
the set of values $\mathcal{T}$. As each record may for example represent a database
system record or a dataset instance or labelled example, we refer to elements of $\mathcal{T}$ as
{\it tuples}.
Reflecting constraints on permissible
database members---\eg representing correlations known publicly and in particular by an 
adversary---databases are elements of some chosen $\mathcal{I}\subseteq\mathcal{T}^n$. 
Importantly,
we do not assume that the data is independent or that it was generated by some
stochastic process.

We define a {\it secret graph} $G=(\mathcal{T},E)$ on the database constituent values,
to be a simple graph with vertex set the tuple values $\mathcal{T}$. The edge set
$E\subseteq\mathcal{T}\times\mathcal{T}$ reflects which value pairs must be kept 
indistinguishable to the adversary.

\begin{definition}[Blowfish policies]
A {\it Blowfish policy} $P=(G, \mathcal{I})$ comprises a secret graph $G=(\mathcal{T},E)$
over database tuple values $\mathcal{T}$ and a (possibly constrained) set of permissible
databases on $n$ tuples, $\mathcal{I}\subseteq\mathcal{T}^n$.
\end{definition}

We next make four preliminary definitions that lift secret
tuple pairs to secret database pairs.
For databases $D,D'\in\mathcal{I}$ the {\it total difference}
is the set of tuples which differ between $D$ and $D'$, in particular it is the
set of triples $(i,u,v)$ which indicate that the $i$th tuples in $D,D'$
are $u$ and $v$ respectively:
\begin{gather*}
    \tdiff(D,D') = \left\{
        (i,u,v)\in[n]\times\mathcal{T}^2    \mid
        u=D_i,v=D'_i,u\neq v
    \right\}
\end{gather*}
The {\it secret difference} between $D$ and $D'$ is the subset of the {\it total difference}
for which $u$ and $v$ are kept secret under Blowfish policy $P$:
\begin{gather*}
    \sdiff(D,D') = \left\{ (i, u, v)\in \tdiff(D,D') \mid
        (u,v)\in E
    \right\}\enspace.
\end{gather*}
\begin{definition}[Minimally secretly different]\label{def:minsecretelydiff}
    A pair of databases $D$ and $D'$
    are {\it secretly different} if they have a non-empty secret difference,
    \ie $\sdiff(D,D')\neq\emptyset$. Two databases $D,D'\in\mathcal{I}$
    are {\it minimally secretly different}
    under Blowfish policy $P$ if both
    \begin{enumerate}[label=(\alph*)]
    \item (secretly different) $\sdiff(D,D') \neq \emptyset$; and\\[-1.0em]
    \item (no closer intermediate database)
    There exists no secretly different $D''\in\mathcal{I}$
    (\ie with $\sdiff(D,D'') \neq\emptyset$) satisfying
    either\\[-1.0em]
        \begin{enumerate}[label=(\roman*)]
            \item (smaller secret difference)
                $\sdiff(D,D'') \subsetneq \sdiff(D,D')$;\footnote{We use $\subsetneq$ and $\supsetneq$ to denote
                proper subset and proper superset respectively.} or
            \item (same secret difference, smaller total difference)
                $\sdiff(D,D'')=\sdiff(D,D')$ and
                $\tdiff(D,D'')\subsetneq \tdiff(D,D')$.
        \end{enumerate}
    \end{enumerate}
\end{definition}
\begin{definition}[Database adjacency graph] \label{def:neighb}
A Blowfish policy $P=(G, \mathcal{I})$ induces a 
{\it database adjacency graph} with vertex set 
$\mathcal{I}$. Two databases $D,D'\in\mathcal{I}$ are adjacent
in this graph---\ie $D \sim D'$---if and only if they are minimally secretly different.
\end{definition}

This definition describes the pairs of databases over which we require a
Blowfish private mechanism to have a
smooth response distribution. In the
differential privacy setting we are concerned with pairs of databases which
differ in a single tuple. The definition of minimally secretly different
generalises this concept for the Blowfish privacy setting where we have
a restricted set of permissible databases $\mathcal{I}$ and secret
values $(u,v)\in E$.

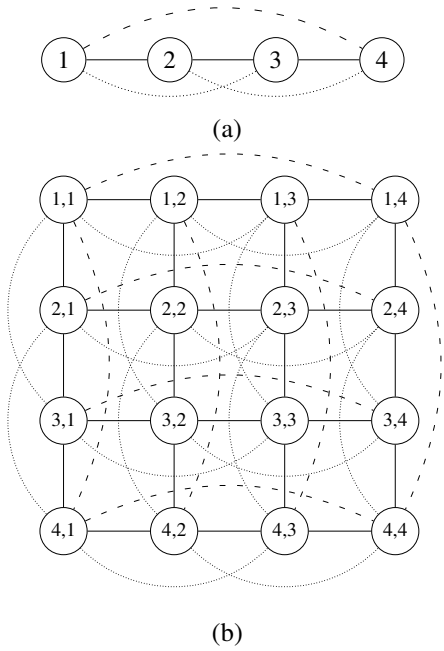
\begin{figure}[!t]
    \begin{minipage}{0.48\textwidth}
    \centering
    \scalebox{0.94}{\input{figures/secrets.tikz}} \\
    (a) \\[1.0em]
    \end{minipage}\hfill
    \begin{minipage}{0.48\textwidth}
    \centering
    \scalebox{0.735}{\input{figures/adjacency.tikz}} \\[0.34em]
    (b)
    \end{minipage}
    \caption{Example~\ref{ex:policy}'s
        (a) distance-threshold secrets graph $G_\theta$ on
        $\mathcal{T}=[4]:=\{1,\ldots,4\}$.
        $\theta=1$ corresponds to only the solid edges, $\theta=2$ also includes
        the dotted edges, and $\theta=3$ includes all edges; and
	(b) the corresponding database adjacency graph for the policy and $n=2$.} %
    \label{fig:eg}
\end{figure}

In the case where the set of permissible databases is unconstrained,
\ie $\mathcal{I}=\mathcal{T}^n$, the database adjacency relationship simplifies
to one more analogous to differential privacy.

\begin{theorem}
    Suppose $\mathcal{I}=\mathcal{T}^n$ and $D,D'\in\mathcal{I}$, then
    $D\sim D'$ if and only if $D$ and $D'$ differ only on a single index
    $i\in[n]$ and the values $D_i$ and $D'_i$ are to be kept secret.
    That is, $D\sim D'$ if and only if $\tdiff(D,D')=\sdiff(D,D')=\{(i,u,v)\}$.
\end{theorem}

\begin{IEEEproof}
    ($\Leftarrow$) Assume, for $i\in[n]$ and $u,v\in\mathcal{T}$ we
    have $\tdiff(D,D')=\sdiff(D,D')=\{(i,u,v)\}$.
    Definition~\ref{def:minsecretelydiff}~(a) holds as $\sdiff(D,D')\neq\emptyset$.
    Definition~\ref{def:minsecretelydiff}~(b) must hold since $\tdiff(D,D')$ and
    $\sdiff(D,D')$ are singleton sets and hence have no non-empty proper subsets.
    
    ($\Rightarrow$) Assume $D\sim D'$, so $\tdiff(D,D') \supseteq \sdiff(D,D') \neq \emptyset$.
    We must show that $|\tdiff(D,D')|=|\sdiff(D,D')|=1$.
    Since $\tdiff(D,D') \neq\emptyset$ there must exist $i\in[n]$,$u,v\in\mathcal{T}$
    such that $(i,u,v)\in \tdiff(D,D')$. Assume for the sake of
    contradiction that there is $j\in[n]\setminus \{i\}$ and $x,y\in\mathcal{T}$
    such that $(j,x,y)\in \tdiff(D,D')$, \ie assume that $|\tdiff(D,D')| > 1$.
    Consider $D''$ which only differs from $D$ at $i$, where $D''_i=v$.
    So $\sdiff(D,D'')=\{(i,u,v)\}\neq\emptyset$ and then either
    $\sdiff(D,D'')\subsetneq \sdiff(D,D')$
    or $\sdiff(D,D'')=\sdiff(D,D'')$ with $\tdiff(D,D'')\subsetneq \tdiff(D,D')$.
    So Definition~\ref{def:minsecretelydiff}~(b) doesn't hold, contradicting
    $D\sim D'$. Hence $|\tdiff(D,D')| \leq 1$. And so since
    $0 < |\sdiff(D,D')| \leq |\tdiff(D,D')| \leq 1$ we have
    that $\tdiff(D,D')=\sdiff(D,D')=\{(i,u,v)\}$.
\end{IEEEproof}

It is clear that we recover the differential privacy adjacency relationship
if $\mathcal{I}=\mathcal{T}^n$ and the secret graph is a clique.

\begin{example}\label{ex:policy}
    An example introduced by He \etal~\cite[Section 3.1]{Blowfish} is the {\it distance threshold}
    secret. For $\mathcal{T}$ with the metric $d$ and some $\theta\in\mathbb{R}$
    the distance threshold secrets graph
    is $G_\theta=(\mathcal{T},E)$, with $(u,v)\in E$ if $d(u,v) \leq \theta$.
    Figures~\ref{fig:eg}(a) and (b) show secret and adjacency graphs respectively for a simple example. Applications of distance threshold secrets include data on age and salary.
\end{example}

\vspace{0.2em}

\begin{definition}[Blowfish privacy]
    Let $\epsilon > 0$ and $P=(G, \mathcal{I})$
    be a policy with induced database adjacency graph $(\mathcal{I},\sim)$.
    A randomised mechanism $\mathcal{K}$ is said to be
    {\it $(\epsilon,P)$-Blowfish private} if, for all $D,D'\in\mathcal{I}$, $D\sim D'$ and all measurable $S\subseteq\operatorname{range}(\mathcal{K})$, mechanism
    $\mathcal{K}$ satisfies
    \begin{equation*}
        \Pr(\mathcal{K}(D)\in S) \leq \exp(\epsilon)\cdot \Pr(\mathcal{K}(D') \in S)\enspace.
    \end{equation*}
\end{definition}

Note that differential privacy is a special case of 
Blowfish privacy where: the secret graph $G$ is a 
complete graph over $\mathcal{T}$ \ie 
$E=\mathcal{T}^2$; permissible datasets are unconstrained
$\mathcal{I}=\mathcal{T}^n$; and as a result,
$\sim$ reduces to the usual neighbouring relation from differential privacy.

\subsection{Quantitative Information Flow} \label{sec:infoflow}

Quantitative information flow~\cite{QuantInfoFlow} models an information-theoretic channel
as a triple $(\mathcal{X}, \mathcal{Z}, K)$. 
Representing channel input and output, $X$ and $Z$ are discrete random
variables (\viz Remark~\ref{rem:discrete}) over the domains
$\mathcal{X} = \left\{ x_1, \ldots, x_{\ell} \right\}$ and
$\mathcal{Z} = \left\{ z_1, \ldots, z_{p} \right\}$ respectively.
$K$ represents the channel matrix 
conditional probabilities $K_{i,j}=\Pr(Z = z_j \mid X = x_i)$.
And if the prior distribution $\pi$ over $X$ is such that $\pi_i=\Pr(X=x_i)$,
then the joint probability distribution over $X$ and $Z$ factors as 
$p(x_i,z_j) = \Pr(X = x_i) \Pr(Z = z_j\mid X = x_i) = \pi_i K_{i,j}$.

The {\it vulnerability} of random variable $X$ is defined by $V(X)=\max_{i\in [\ell]}
\Pr(X=x_i)$, representing the worst-case probability that an adversary can correctly
guess the value of $X$ in a single try. Similarly, the {\it conditional vulnerability}
representing the probability of an adversary correctly guessing $X$ in a single try
after observing $Z$, is defined by $V(X | Z) =  \sum_{j\in [p]}
\max_{i\in [\ell]} \Pr(X=x_i)\Pr(Z=z_j\mid X=x_i)$.

Measured as information,  vulnerability is equivalent to
the {\it min-entropy} $H_\infty (X) = -\log V(X)$ of $X$,
and the {\it conditional min-entropy} $H_\infty (X | Z)
= -\log V(X | Z)$ of $X$ given $Z$. %
                    We use the notation $H^{K}_\infty(X|Z)$ to refer to the min-entropy
                    $H_\infty(X|Z)$ for the channel matrix $K$ when the channel matrix
                    in question is not clear from the context. %
{\it Information leakage} (or {\it min-entropy leakage}) is the difference between
the min-entropy before and after observing the output $Z$, \ie $I_\infty(X;Z)=
H_\infty(X) - H_\infty(X | Z)$.
We will make use of a simplification of the min-entropy of channel matrices under uniform prior.

\begin{lemma} \label{lem:Hofmatrix}
    Let $(\mathcal{X},\mathcal{Z},K)$ be an information-theoretic channel, with
    $X,Z$ random variables over domains $\mathcal{X}$ and
    $\mathcal{Z}$ respectively. $K$ is the $\ell\times p$ channel
    matrix. If $X$ has the uniform distribution over $\mathcal{X}$ then,
        $H_\infty(X|Z) = -\log \frac{1}{\ell} \sum_{j=1}^{p} \max_i K_{i,j}$,
    \ie the information leakage of the channel is equal to the sum of 
    the column maxima of $K$.
\end{lemma}

\subsection{Differential Privacy Implies Bounded Information Leakage}

Alvim \etal~\cite{MSAlvim2} consider a differentially-private mechanism $\mathcal{K}$ as an
information-theoretic channel $(\mathcal{X},\mathcal{Z},K)$ with 
$\mathcal{X}= \mathcal{I}$ the set of permissible databases,
$\mathcal{Z}=\operatorname{range}(\mathcal{K})$ the mechanism's response space,
and $K$ the $\ell \times p$ channel matrix with $K_{i,j}=\Pr(\mathcal{K}(x_i) = z_j)$.
They established that the differential privacy of $\mathcal{K}$
implies an upper bound on the information leakage for the corresponding channel. They
also demonstrate that this implication does not go the other way: a channel with known
information leakage does not necessarily satisfy $\epsilon$-differential privacy for any
$\epsilon$.

Note that when discussing channel matrices we will often refer to
elements of the input and output sets by their indices, \eg writing $x_i \sim x_h$ as
$i \sim h$. 
A release mechanism $\mathcal{K}$ with corresponding channel matrix $K$ being
$(\epsilon,P)$-Blowfish private is equivalent to the statement that, for all $i,h\in[\ell]$
and all $j\in[p]$ such that 
$i\sim h$,
\begin{equation} \label{eq:channelblowfish}
	\exp(-\epsilon) \leq \frac{K_{i,j}}{K_{h,j}} \leq \exp(\epsilon)\enspace.
\end{equation}

\begin{remark}\label{rem:discrete}
Like Alvim \etal~\cite{MSAlvim2}, we assume channels with discrete input
and output spaces which correspond to discrete data and responses.
Rounding due to finite precision in floating-point implementations of
private mechanisms can cause low-probability responses to become
zero-probability~\cite{MironovFloatingPoint}, violating differential/Blowfish privacy.
It is therefore 
regarded best practice that privacy analysis of mechanisms 
require discrete response distributions~\cite[Remark~2.1]{DworkBook}.
We assume suitably discretised distributions.
\end{remark}

\section{Main Results}

In this section we present and discuss Theorem~\ref{thm:minent}
and Main~Theorem~\ref{thm:infleak} which bound the
min-entropy and information leakage of Blowfish-private
mechanisms. Proofs for these results are given in
Section~\ref{sec:symmetrisation}.

Maximum information leakage 
is attained for a uniform prior over input
$\mathcal{X}$~\cite{braun2009leakage}.
As a result, we can assume a uniform prior
in order to derive a general upper bound on information leakage
for $\mathcal{K}$,
holding when the random variables $X$ and $Z$ have any distribution over $\mathcal{X}$
and $\mathcal{Z}$.

\begin{theorem}[Min-entropy of Blowfish-private mechanisms] \label{thm:minent}
    Let $\epsilon > 0$ and $P$ be
    a Blowfish policy. Let $(\mathcal{X},\mathcal{Z},K)$ be the channel which
    corresponds to a mechanism $\mathcal{K}$ satisfying
    $(\epsilon,P)$-Blowfish privacy. If $X$ has the uniform distribution then,
    \begin{equation*}
        H_\infty(X | Z) \geq -\log \left(\frac{1}{\ell} \sum_{t=1}^q \exp\left(\epsilon d_t\right)\right)\enspace,
    \end{equation*}
    where $\ell = |\mathcal{X}|$,
    $q$ is the number of connected components of database adjacency graph
    $(\mathcal{X},\sim)$ and, for $t\in[q]$, $d_t$ is the $t^\text{th}$ connected
    component's diameter, \ie the maximal shortest-path distance between any pair of vertices in the
    component.
\end{theorem}

Motivating examples of Blowfish adjacency graphs from the literature~\cite{Blowfish,haney2015design}
are frequently connected or have $q\ll \ell$.

\begin{example}
Revisiting the $G_\theta$ secret graph of Example~\ref{ex:policy}, the induced adjacency graph $(\mathcal{X},\sim)$
is connected unless $\mathcal{T}$ contains consecutive values $u,v$ such that
$d(u,v) > \theta$.
\end{example}

\begin{maintheorem}[Information leakage of Blowfish-private mechanisms] \label{thm:infleak}
    Let $\epsilon > 0$ and $P$ be
    a Blowfish policy. Let $(\mathcal{X},\mathcal{Z},K)$ be the channel which
    corresponds to a $(\epsilon,P)$-Blowfish-private mechanism $\mathcal{K}$.
    Then there is an upper bound on the information
    leakage of $K$,
    \begin{equation} \label{eq:infleak}
        I_\infty(X;Z) \leq \log\left( \sum_{t=1}^q \exp\left(\epsilon d_t\right)\right)\enspace,
    \end{equation}
    where $q$ is the number of connected components of
    $(\mathcal{X},\sim)$ and $d_t$ is the diameter of the $t^\text{th}$ connected
    component for $t\in[q]$.
    Note that this result holds for all prior distributions on $\mathcal{X}$.
\end{maintheorem}

Recall here that
$(\mathcal{X},\sim)$ is the database adjacency graph,
with $\mathcal{X}=\mathcal{I}$. 
In the case that $(\mathcal{X},\sim)$ is connected, (\ref{eq:infleak})
simplifies to $I_\infty(X;Z)\leq \epsilon d$, where $d=\operatorname{Diam}(\mathcal{X},\sim)$.
As expected, increasing the level of Blowfish privacy
(by decreasing $\epsilon$) pushes down the bound 
on information leakage.

As discussed in Example~\ref{eg:cyclecomplete} the differential
privacy case corresponds to $\mathcal{I}=\mathcal{T}^n$
and a complete secret graph on $\mathcal{T}$. Hence
the database adjacency graph $(\mathcal{X}, \sim)$
is connected with diameter $n$, and the bound simplifies
to $\epsilon n$. For an unconstrained set of databases
$\mathcal{I}=\mathcal{T}^n$ and a connected secret graph,
the diameter of $(\mathcal{X},\sim)$ is given by
$n$ times the diameter of the secret graph.
A larger diameter of the secret graph, and hence a larger
diameter for the database adjacency graph in this case,
arises when there are fewer pairs of values to be kept
secret.

\begin{figure}[b]
    \centering
    \scalebox{0.75}{\input{figures/leakageplot.tikz}}
    \caption{
            Upper bound on Example~\ref{ex:policy}'s information leakage
            for $\mathcal{T}=[4]$ as $n$ is varied, $\epsilon$ held
            constant.%
        \label{fig:bound}}
\end{figure}
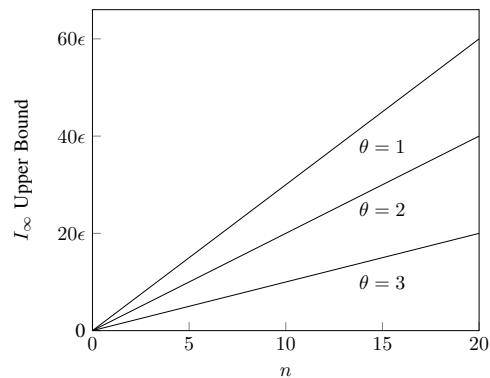

Fewer secret value pairs allows our mechanism
to attain the same level of Blowfish privacy (\ie the same $\epsilon$)
while adding less perturbation to the response.
In other words, when we are concerned about revealing differences
between a smaller set of values, there is a smaller
set of responses over which the channel's probability distribution
must be smooth.

While Blowfish privacy measures only the
level of privacy on values to be kept indistinguishable,
min-entropy and related privacy loss measures do not
encapsulate such fine-grained policies. That is, revealing information
about ``secrets'' and ``non-secrets'' impacts
the information leakage equally.
The increase in Main Theorem~\ref{thm:infleak}'s bound
corresponding to fewer pairs of secrets while holding $\epsilon$
fixed is consistent with this difference between the definitions.

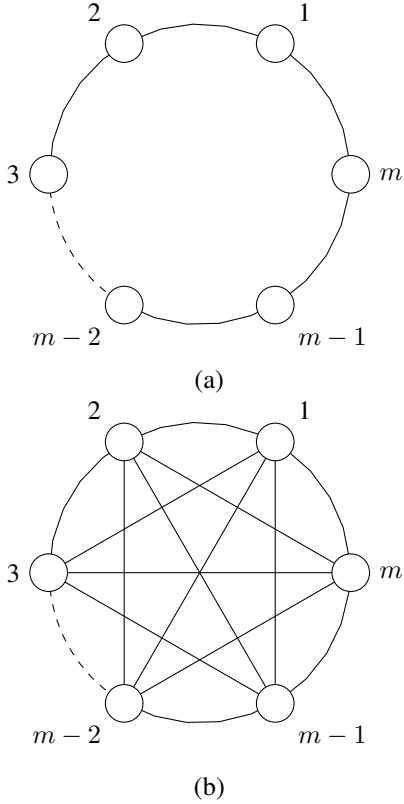
\begin{figure}[!t]
    \begin{minipage}{0.48\textwidth}
    \centering
    \scalebox{1}{\input{figures/cyclesecrets.tikz}} \\
    (a) \\[1.0em]
    \end{minipage}\hfill
    \begin{minipage}{0.48\textwidth}
    \centering
    \scalebox{1}{\input{figures/completesecrets.tikz}} \\[0.34em]
    (b)
    \end{minipage}
    \caption{Example~\ref{eg:cyclecomplete} introduces
    (a) the cycle secret graph on $\mathcal{T}=[m]$ and
    (b) the complete secret graph on $\mathcal{T}=[m]$.
    The diameter of the secret graph is $\left\lfloor \frac{m}{2}\right\rfloor$
    for (a) and $1$ for (b).
    }
    \label{fig:egsecretgraphs}
\end{figure}

\begin{example}
    Again revisiting Example~\ref{ex:policy}'s $G_\theta$ secret graph,
    Figure~\ref{fig:bound} plots the relationship between $n$ and
    our bound on information leakage for $\theta\in\{1,2,3\}$.
    Increasing $\theta$ corresponds to holding more pairs of values secret, so the
    slower growth under larger $\theta$ is as expected.
\end{example}

\begin{example} \label{eg:cyclecomplete}
    Suppose $|\mathcal{T}|=m$ and $\mathcal{I}=\mathcal{T}^n$.
    Contrast the case where the secret graph is
    (a) a cycle (\ie $(u, v)\in E$ when
    $u \equiv v + 1 \mod m$ or $u \equiv v - 1 \mod m$)
    with (b) a complete secret graph.
    
    A cycle graph may arise when values correspond
    to angles, latitudes, or times of the day. Blowfish privacy
    in the complete secret
    graph case coincides with differential privacy.
    
    For (a) the diameter of the secret graph is
    $\left\lfloor \frac{m}{2} \right\rfloor$, and hence the
    diameter of the induced database adjacency
    graph $(\mathcal{X},\sim_\text{(a)})$
    is $n\left\lfloor \frac{m}{2}\right\rfloor$. For (b)
    the diameter of the complete secret graph
    is $1$, and so the database adjacency graph
    $(\mathcal{X},\sim_\text{(b)})$ has diameter $n$.
    The upper bounds on information leakage
    are (a) $\epsilon n \left\lfloor \frac{m}{2} \right\rfloor$
    and (b) $\epsilon n$.
\end{example}

Under differential privacy (a special case of Blowfish), the adjacency
graph exhibits distance regularity and vertex transitivity.  
Previous work has focused only on differential privacy and has therefore
made strong assumptions in the form of this graph symmetry~\cite{MSAlvim2}. 
The challenge in proving our main results is that we may make no symmetry 
assumptions (\viz Theorem~\ref{thm:arbitrarygraph}). 
We therefore symmetrise by organising the graph into vertex-transitive
automorphism orbits.

\begin{theorem} \label{thm:arbitrarygraph}
    Let $A=(V_A, E_A)$ be an arbitrary undirected graph. Then there exists a Blowfish
    policy $P=(G,\mathcal{I})$ such that the induced database adjacency graph
    $(\mathcal{I},\sim)=(V_A,E_A)$.
\end{theorem}

\begin{IEEEproof}
    Let the secret graph $G=(\mathcal{T},E)$ where $\mathcal{T}=V_A$ and
    $E=E_A$, \ie the set of tuples coincides with the vertices in $A$ and the
    secret pairs of values correspond to adjacent vertices in $A$.
    Let the set of permissible databases be all databases with one tuple,
    so $\mathcal{I}=\mathcal{T}^1$.
    
    Now we must show that two databases $D=(u),D'=(v)\in\mathcal{I}$
    are adjacent in the database adjacency graph (\ie $D\sim D'$) if and only if
    $u$ and $v$ are adjacent in $A$ (\ie $(u,v)\in E_A$). That is, we must show
    that $D$ and $D'$ are minimally secretly different (Definition~\ref{def:minsecretelydiff})
    iff $(u,v)\in E_A$.
    
    We have three cases:
    (a) $u=v$, (b) $u\neq v$ and $(u,v)\not\in E_A$, or (c) $u\neq v$
    and $(u,v)\in E_A$.
    In case (a) we have $\sdiff(D,D')\subseteq \tdiff(D,D') = \emptyset$
    and thus Definition~\ref{def:minsecretelydiff}.a does not hold, so $D$
    and $D'$ are not neighbouring databases. Also since $u=v$, $(u,v)\not\in E_A$.
    In case (b) we have $\tdiff(D,D')=\{(1,u,v)\}$, but since $(u,v)\not\in E_A$
    we have $(u,v)\not\in E$ and so $\sdiff(D,D')=\emptyset$.
    Again, Definition~\ref{def:minsecretelydiff}.a does not hold and so $D$
    and $D'$ are not neighbouring databases.
    
    In case (c) we have $\tdiff(D,D')=\{(1,u,v)\}$. Since $(u,v)\in E_A$
    we have $(u,v)\in E$ and so $\sdiff(D,D')=\{(1,u,v)\}$ as well.
    Now Definition~\ref{def:minsecretelydiff}.a is satisfied. To show $D\sim D'$
    we need to demonstrate that there does not exist $D''$ with 
    $\sdiff(D,D'')\neq\emptyset$ satisfying
    Definition~\ref{def:minsecretelydiff}.a.i or
    Definition~\ref{def:minsecretelydiff}.a.ii. Suppose there exists
    $D''\in\mathcal{I}$ with $\sdiff(D,D'')\neq\emptyset$. The maximum size of
    $\tdiff(D,D'')$ is 1 since $\mathcal{I}$ is the set of databases with
    1 tuple, so $|\tdiff(D,D'')| \leq 1$. Also since
    $\sdiff(D,D'')\neq\emptyset$, $1\leq |\sdiff(D,D'')|$.
    Combining these properties, along with the fact that $\sdiff(D,D'')\subseteq
    \tdiff(D,D'')$ we have
    $1\leq |\sdiff(D,D'')|\leq |\tdiff(D,D'')| \leq 1$.
    So $\sdiff(D,D'')$ must not be a proper subset of $\sdiff(D,D')$
    and $\tdiff(D,D'')$ must not be a proper subset of $\tdiff(D,D')$;
    thus neither Definition~\ref{def:minsecretelydiff}.a.i not
    Definition~\ref{def:minsecretelydiff}.a.ii are satisfied. So $D$ and $D'$
    are not adjacent in $(\mathcal{I},\sim)$.
    
    So, in all possible cases $D,D'\in\mathcal{I}$ are minimally secretly different---and
    hence $D\sim D'$ in $(\mathcal{I},\sim)$---if and only if $(u,v)\in E_A$. So
    the policy $P$ induces a database adjacency graph $(\mathcal{I},\sim)$
    which coincides with the arbitrary graph $A=(V_A,E_A)$.
\end{IEEEproof}

\begin{figure*}[!h]
    \begin{equation*}
    K^{(\delta)}
    = \frac{1}{4+2\delta}
    \begin{pmatrix}
    \begin{matrix}
        1+\delta & 1+\delta & 1 & 1 \\
        1 & 1+\delta & 1+\delta & 1 \\
        1 & 1 & 1+\delta & 1+\delta \\
        1+\delta & 1 & 1 & 1+\delta \\
        \end{matrix} & & & \begin{matrix} \phantom{2} \cdots \phantom{2\delta} & \phantom{2} 0 \phantom{2\delta} \\ & \phantom{2} \vdots \phantom{2\delta} \\ & \\ & \\ \end{matrix} \\
        & \begin{matrix}
            2+2\delta & 2 \\
            2 & 2+2\delta \\
            \end{matrix} & & \\
                \begin{matrix} \phantom{1} \vdots \phantom{\delta} & \phantom{1+\delta} & \phantom{1+\delta} & \phantom{1+\delta} \\ \end{matrix} & & \ddots & \\
                    \begin{matrix} 0 \phantom{2\delta} & \cdots & \phantom{1+\delta} & \phantom{1+\delta} \\ \end{matrix} & & & \begin{matrix}
            2+2\delta & 2 \\
            2 & 2+2\delta \\
        \end{matrix} \\
    \end{pmatrix}\enspace.
    \end{equation*}
    \caption{Family of block diagonal channel matrices (parametrised by real $\delta>0$) used in the construction of Theorem~\ref{thm:sharp}.}
    \label{fig:thm-sharp}
    \end{figure*}

\begin{figure}[b]
    \centering
    \scalebox{0.9}{\input{figures/sharpness.tikz}}
        \caption{
            The database adjacency graph used in Theorem~\ref{thm:sharp}'s construction 
            demonstrating tightness of our main bound on information leakage.%
        \label{fig:sharpness}}
\end{figure}

We next construct a family of channel matrices and an adjacency graph that asymptotically meet our bound with equality, and for which previous bounds~\cite{MSAlvim1, MSAlvim3, MSAlvim2} do not hold.
This demonstrates that the bound is tight in the limit. In
particular, we describe an scenario for which a smaller upper bound
would not hold.
\begin{theorem}
\label{thm:sharp} 
There exists a family of mechanisms $\mathcal{K}^{(\delta)}$, for $\delta>0$, and a Blowfish policy $P$, such that the Main Theorem~\ref{thm:infleak} upper bound on information leakage is equal to the information leakage, asymptotically. Namely,
	\begin{equation*}
		\lim_{\delta\downarrow 0}\frac{\log\left(\sum_{t=1}^q \exp(\epsilon(\delta)\cdot d_t)\right)}{\mathcal{I}_\infty\left(X;K^{(\delta)}(X)\right)} = 1\enspace,
	\end{equation*}
	where $K^{(\delta)}$ denotes the channel matrix for mechanism $\mathcal{K}^{(\delta)}$ and $K^{(\delta)}(X)$ denotes its output random variable $Z$, $\epsilon(\delta)$ represents the Blowfish privacy level of $\mathcal{K}^{(\delta)}$ with respect to policy $P$, $q$ and $d_1,\ldots,d_q$ are the number of connected components of $(\mathcal{I},\sim_P)$ and the corresponding component diameters. In particular, the policy $P$'s induced adjacency graph $(\mathcal{I},\sim_P)$ is neither vertex transitive nor distance regular.
\end{theorem}

\begin{IEEEproof}
Consider a fixed integer $n>1$, and define the undirected graph $(\mathcal{X},E_X)$ with nodes $x_1,\ldots,x_{2n+2}$ as shown in Figure~\ref{fig:sharpness}: one complete connected component $\{x_1,\ldots,x_4\}$, and $n-1$ complete connected components $\{x_5,x_6\},\ldots,\{x_{2n+1},x_{2n+2}\}$, for a total of $q=n$ connected components. Note that: component $t\in [n]$ has diameter $d_t=1$; the graph is not regular and so cannot be vertex transitive; and because there are $2$ nodes at distance one from connected nodes $x_1, x_2$ but no nodes at distance one from connected nodes $x_5, x_6$, that the graph is not distance regular. By Theorem~\ref{thm:arbitrarygraph}, there exists a Blowfish policy $P=(G,\mathcal{I})$ such that the database adjacency graph $(\mathcal{I},\sim)=(\mathcal{X},E_X)$,
where the permissible databases $\mathcal{I}$ coincide with the elements of $\mathcal{X}$.
As it is the database adjacency graph that directly impacts our bound on information leakage, we will not make further reference to details of $P$.

For real $\delta>0$, consider the block diagonal channel matrix $K^{(\delta)}$, shown in Figure~\ref{fig:thm-sharp}, with input variable $X$ uniformly distributed on the vertex set of our constructed graph $\mathcal{X}=\{x_1,\ldots,x_{2n+2}\}$, and output variable $Z^{(\delta)}$ on finite space $\mathcal{Z}$ of cardinality $2n+2$.

Each row of $K^{(\delta)}$ is normalised by dividing by constant $4+2\delta$. By construction of the block structure and common normalising constants, the maximum ratio of any two elements within a column, between rows $x_i\sim x_j$ is simply $1+\delta$ and therefore the corresponding mechanism $\mathcal{K}^{(\delta)}$ preserves $(\epsilon(\delta),P)$-Blowfish privacy for $\epsilon(\delta) := \log(1+\delta)$, independent of free parameter $n$.%

        Since $X$ is uniformly distributed over $\mathcal{X}$ the information leakage
        of $\mathcal{K}^{(\delta)}$ is given by the log of the sum of column maxima,
        \begin{align*}
            & I_\infty(X;K^{(\delta)}(X)) \\
                =& H_\infty(X) - H_\infty(X|Z^{(\delta)}) \\
                =& \log\left(\frac{V(X|Z^{(\delta)})}{V(X)}\right) \\
                =& \log\left(
                    \frac{
                        \sum_{j=1}^{2n+2}
                        \max_{i\in[2n+2]}
                        \Pr(X=x_i)C_{ij}
                    }{
                        \max_{i\in[2n+2]} \Pr(X=x_i)
                    }
                \right)
                \\
                =& \log\left(
                    \frac{
                        \frac{1}{2n+2}\sum_{j=1}^{2n+2}
                        \max_{i\in[2n+2]}K^{(\delta)}_{ij}
                    }{
                        \frac{1}{2n+2}
                    }
                \right) \\
                =& \log\left(\frac{
                    4(1+\delta) + (2n - 2)(2+2\delta)
                }{4+2\delta}\right)\enspace.
        \end{align*}
    
    where $C_{ij}$ denotes $\Pr(Z^{(\delta)}=z_j|X=x_i)$.
	Putting these calculations together, noting that the limit of the ratio
	of logs is the ratio of logs of the limits since $\log$ is a continuous function and
	the denominator is a positive constant in the limit,
	\begin{align*}
	    & \lim_{\delta\downarrow 0}
		\frac{\log\left(\sum_{t=1}^q \exp(\epsilon(\delta)\cdot d_t)\right)}{\mathcal{I}_\infty\left(X;K^{(\delta)}(X)\right)}  \\
		=&
	    \lim_{\delta\downarrow 0}
		\frac{\log(n(1+\delta))}{\log\left(\frac{4n(1+\delta)}{4+2\delta}\right)} 
		= 
		\frac{\log(n)}{\log(n)} 
		= 1\enspace.
	\end{align*}
	
    As such we have constructed a family of mechanisms $\mathcal{K}^{(\delta)}$
    which are $(\epsilon(\delta),P)$-Blowfish private for $\epsilon(\delta):=1+\delta$
    and Blowfish policy $P$ such that the upper bound from Main~Theorem~\ref{thm:infleak}
    is attained in the limit, as $\delta\downarrow0$.
\end{IEEEproof}

\section{Graph Symmetrisation}\label{sec:symmetrisation}

To prove Theorems~\ref{thm:minent} and~\ref{thm:infleak} 
we perform matrix operations which maintain the $(\epsilon,P)$-Blowfish privacy and information leakage of the
channel $\mathcal{K}$. %

The first of these transformations (Lemma~\ref{lem:kprime}) takes the $\ell \times p$ channel matrix $K$ to an
$\ell \times \ell$ channel matrix $K'$ such that each column attains its maximum value
in the diagonal.
This matrix $K'$ satisfies $(\epsilon,P)$-Blowfish privacy and attains the same
information leakage as $K$.

Second (\viz Lemma~\ref{lem:kprimeprime}), this $K'$ is transformed into an $\ell \times \ell$ channel matrix $K''$, also
satisfying $(\epsilon,P)$-Blowfish privacy and maintaining the same information leakage.
Additionally all diagonal elements of $K''$ which are in the same orbit of some automorphism group $\Gamma$
over the adjacency graph are equal: for all if $i,h\in[\ell]$ members of the same $\Gamma$-orbit, 
$K''_{i,i}=K''_{h,h}$.
A key property of these orbits is that they are vertex transitive.

In the specific case where $X$ has the uniform distribution over $\mathcal{X}$, the properties
of the partitions of $K''$ allow us to find a lower bound for min-entropy. Since
information leakage achieves its maximum over the uniform distribution this allows us
to bound information leakage for arbitrary priors over $\mathcal{X}$.

\paragraph{Abstract algebra basics.}
Before detailing our results, we 
list group-theoretic notation required for our proofs.  
We focus on the database adjacency graph 
$(\mathcal{X},\sim)$ induced by chosen Blowfish policy $P$. 
Note that the neighbouring relation $\sim$ imposes no
restrictions on the graph: Theorem~\ref{thm:arbitrarygraph} 
demonstrates that any simple graph can be induced by $P$. 
In particular, $(\mathcal{X},\sim)$ need not be vertex transitive nor distance regular
unlike adjacency graphs under differential privacy~\cite{MSAlvim2}.
We use $d(x_i,x_j)$ to denote the geodesic distance between $x_i$ and $x_j$ in $(\mathcal{X},\sim)$,
\ie the number of edges in a shortest path connecting $x_i$ and $x_j$.

We refer to the {\it (full) automorphism group} of $(\mathcal{X},\sim)$ by
$\Aut(\mathcal{X},\sim)$, and consider $\Gamma\subseteq\Aut(\mathcal{X},\sim)$
to be an {\it automorphism (sub)group} of $(\mathcal{X},\sim)$.
For $u\in\mathcal{X}$, denote the {\it stabiliser} of $u$ in $\Gamma$ by
$\Gamma_u=\{\sigma\in\Gamma \:|\: \sigma(u) = u\}$. Additionally, denote the {\it $\Gamma$-orbit}
of $u$ in $\mathcal{X}$ by $\Gamma(u)=\{\gamma(u) \:|\: \gamma\in\Gamma\}\subseteq \mathcal{X}$.

If $\Gamma(u)=\mathcal{X}$ for some (and hence all) $u\in\mathcal{X}$ then $\Gamma$ is said
to be {\it transitive} on $\mathcal{X}$ and $\mathcal{X}$ is said to be {\it
$\Gamma$-vertex transitive}. We say that $(\mathcal{X},\sim)$ is {\it vertex transitive} if
it is $\Aut(\mathcal{X},\sim)$-vertex transitive.

For $u,v\in\mathcal{X}$ the following notation is introduced in \cite[p.28]{MSAlvim2} to indicate the
set of automorphisms in $\Gamma$ taking $u$ to $v$, \ie $\Gamma_{u\rightarrow v}
=\{\sigma\in\Gamma \:|\: \sigma(u)=v\}$. Note that $\Gamma_{u\rightarrow v}$ is not a group
unless $u=v$, in which case $\Gamma_{u\rightarrow v}=\Gamma_{u\rightarrow u}=\Gamma_u$.

Additional graph-theoretic results used in the proof of Lemma~\ref{lem:kprimeprime}
are introduced next.

\subsection{Technical Symmetrisation Lemmata}

Lemmata~\ref{lem:leftcoset}--\ref{lem:stabsize}
establish group-theoretic facts about the
automorphism groups of undirected graphs that are used in
Section~\ref{subsec:matrixtransforms}
where we prove that transformations of the channel matrix have
well understood effects on the level of Blowfish-privacy
and conditional min-entropy.
Lemmata~\ref{lem:autsizes}~and~\ref{lem:stabsize}
rely on the orbit-stabiliser theorem, a textbook result
in group theory (\eg see
Dixon and Mortimer~\cite[Theorem 1.4A]{DixonPermutations}).
\begin{lemma} \label{lem:leftcoset}
    Let $(\mathcal{X},\sim)$ be an undirected graph. Let $u,v\in\mathcal{X}$.
    Let $\Gamma\in\Aut(\mathcal{X},\sim)$. Let $\sigma\in\Gamma_{u\rightarrow v}$.
    Let $\sigma\Gamma_u$ denote the left coset $\{\sigma\circ\gamma \:|\: \gamma\in\Gamma_u\}$.
    Then,
    \begin{equation*}
        \Gamma_{u\rightarrow v} = \sigma\Gamma_u\enspace.
    \end{equation*}
\end{lemma}

\begin{IEEEproof}
    For $\gamma\in\Gamma_u$ we have $\sigma\circ\gamma(u) =
    \sigma(\gamma(u))=\sigma(u)=v$, so $\sigma\circ\gamma\in\Gamma_{u\rightarrow v}$, so
    $\sigma\Gamma_u\subseteq \Gamma_{u\rightarrow v}$.
    
    Conversely, for $\gamma\in\Gamma_{u\rightarrow v}$ we have $\gamma(u)=v=\sigma(u)$ and
    $u=\sigma^{-1}(v)$. So $\sigma^{-1}\circ\gamma(u)=\sigma^{-1}(\gamma(u))=\sigma^{-1}(v)=u$,
    and as such $\sigma^{-1}\circ\gamma \in \Gamma_u$. Hence $\sigma\circ\sigma^{-1}\circ\gamma
    =\gamma\in\sigma\Gamma_u$. So $\Gamma_{u\rightarrow v}\subseteq \sigma\Gamma_u$.
    
    Since $\sigma\Gamma_u\subseteq \Gamma_{u\rightarrow v}$ and $\Gamma_{u\rightarrow v}\subseteq
    \sigma\Gamma_u$, we have shown that $\Gamma_{u\rightarrow v} = \sigma\Gamma_u$.
\end{IEEEproof}

\begin{lemma} \label{lem:rightcoset}
    Let $(\mathcal{X},\sim)$ be an undirected graph. Let $u,v\in\mathcal{X}$.
    Let $\Gamma\in\Aut(\mathcal{X},\sim)$. Let $\sigma\in\Gamma_{u\rightarrow v}$.
    Let $\Gamma_v\sigma$ denote the right coset $\{\gamma\circ\sigma \:|\: \gamma\in\Gamma_v\}$.
    Then,
    \begin{equation*}
        \Gamma_{u\rightarrow v} = \Gamma_v\sigma\enspace.
    \end{equation*}
\end{lemma}

\begin{IEEEproof}
    For $\gamma\in\Gamma_v$ we have $\gamma(\sigma(u))=\gamma(v)=v$, so $\gamma\circ\sigma \in
    \Gamma_v$, so $\Gamma_v\sigma\subseteq \Gamma_{u\rightarrow v}$.
    
    Consider $\gamma\in\Gamma_{u\rightarrow v}$. We have $\gamma(u)=v=\sigma(u)$, and as such
    $\gamma\circ\sigma^{-1}(v)=\gamma(u)=v$, so $\gamma\circ\sigma^{-1}\in\Gamma_v$
    and $\gamma\circ\sigma^{-1}\sigma=\gamma\in\Gamma_v\sigma$. So $\Gamma_{u\rightarrow v}
    \subseteq\Gamma_v\sigma$.
    
    Since $\Gamma_v\sigma\subseteq \Gamma_{u\rightarrow v}$ and $\Gamma_{u\rightarrow v}
    \subseteq\Gamma_v\sigma$, we have shown that $\Gamma_{u\rightarrow v}=\Gamma_v\sigma$.
\end{IEEEproof}

\begin{lemma} \label{lem:autsizes}
    Let $(\mathcal{X},\sim)$ be an undirected graph. Let $u,v\in\mathcal{X}$.
    Let $\Gamma\in\Aut(\mathcal{X},\sim)$. Then,
    \begin{equation*}
        |\Gamma_{u\rightarrow v}| = |\Gamma_u| = \frac{|\Gamma|}{|\Gamma(u)|}
            = \frac{|\Gamma|}{|\Gamma(v)|} = |\Gamma_v|\enspace.
    \end{equation*}
\end{lemma}

\begin{IEEEproof}
    We start by citing the {\it orbit-stabiliser theorem}. For any $u\in\mathcal{X}$
    we have
    \begin{equation*}
        |\Gamma_u| = \frac{|\Gamma|}{|\Gamma(u)|}\enspace.
    \end{equation*}
    From Lemma \ref{lem:leftcoset} we have, $|\Gamma_{u\rightarrow v}|=|\sigma\Gamma_u|=|\Gamma_u|$.
    Similarly, from Lemma \ref{lem:rightcoset} we have
    $|\Gamma_{u\rightarrow v}|=|\Gamma_v\sigma|=|\Gamma_v|$. So applying the orbit-stabiliser
    theorem, we have
    \begin{gather*}
	    |\Gamma_{u\rightarrow v}| = |\Gamma_u| = \frac{|\Gamma|}{|\Gamma(u)|}\enspace\mbox{and}\enspace 
	    |\Gamma_{u\rightarrow v}| = |\Gamma_v| = \frac{|\Gamma|}{|\Gamma(v)|}\enspace,
    \end{gather*}
	establishing the result.
\end{IEEEproof}

\begin{lemma} \label{lem:stabsize}
    Let $(\mathcal{X},\sim)$ be an undirected graph. Let $u,v\in\mathcal{X}$.
    Let $\Gamma\in\Aut(\mathcal{X},\sim)$. If $\Gamma_{u\rightarrow v}\neq\emptyset$
    then,
    \begin{equation*}
        |\Gamma_u| = |\Gamma_v|\enspace.
    \end{equation*}
\end{lemma}

\begin{IEEEproof}
    Let $\sigma\in\Gamma_{u\rightarrow v}$. Let $\gamma\in\Gamma_u$.
    So $\gamma(u)=u$, and $\sigma(\gamma(u))=\sigma(u)=v$. Also $\sigma^{-1}(v)=u$,
    so $\sigma\circ\gamma\circ\sigma^{-1}(v)=\sigma(\gamma(\sigma^{-1}(v)))=\sigma(\gamma(u))=v$.
    So for any $\sigma\in\Gamma_{u\rightarrow v}$ and $\gamma\in\Gamma_u$
    we have $\sigma\circ\gamma\circ\sigma^{-1}\in\Gamma_v$. So
    $\sigma\Gamma_u\sigma^{-1}\subseteq\Gamma_v$.
    
    Thus we have $|\sigma\Gamma_u\sigma^{-1}|=|\Gamma_u|\leq|\Gamma_v|$. Similarly,
    by exchanging the roles of $u$ and $v$ above, we get that $|\Gamma_v|\leq|\Gamma_u|$.
    Since $|\Gamma_u|\leq|\Gamma_v|$ and $|\Gamma_v|\leq|\Gamma_u|$ then $|\Gamma_u| = |\Gamma_v|$.
\end{IEEEproof}

\subsection{Channel Matrix Transformations}\label{subsec:matrixtransforms}

We now develop the matrix transformations discussed
in the sketch above. First we transform channel matrix $K$ to 
attain column maxima along the diagonal.

\begin{lemma} \label{lem:kprime}
    Let $K$ be an $\ell \times p$ channel matrix such that $K$ satisfies $(\epsilon,P)$-Blowfish
    privacy. Then there exists an $\ell \times \ell$ matrix
    $K'$ such that:
    \begin{enumerate}[label=(\alph*)]
            \item $K'$ is a channel matrix, \ie $K'_{i,j}\in[0,1]$ and $\sum_{h=1}^{\ell}K'_{i,h}=1$
                for all $i,j\in[\ell]$;
            \item Each column $j\in[\ell]$ has a maximum in the diagonal, 
                $K'_{j,j}=\max_{i\in [\ell]}K'_{i,j}$;
            \item $K'$ satisfies $(\epsilon, P)$-Blowfish privacy, 
                $K'_{i,j} \leq e^{\epsilon} K'_{h,j}$ for all $i,j,h\in[\ell]$
                such that $i\sim h$; and
            \item
            If $X$ has the
                uniform distribution over $\mathcal{X}$ then the conditional min-entropies for channel matrices $K,K'$ are equal, \ie $H^{K'}_\infty(X|Z)=H^{K}_\infty(X|Z)$.
    \end{enumerate}
\end{lemma}

\begin{IEEEproof}
    First we assume that $K$ is an $\ell \times p$ matrix, with $\ell \leq p$. This assumption
    is without loss of generality as we can append all-zero columns until this condition is satisfied,
    corresponding to augmenting the output set $\mathcal{Z}$ with elements of
    probability 0. That is, $\mathcal{Z}' = \mathcal{Z} \cup \{z_{p+1}, z_{p+2}, \ldots, z_\ell\}$ holds
	with $\Pr(Z=z_i)=0$ for all $i\in$  $\{p+1,\ldots,\ell\}$.

    From here the proof of \cite[Lemma 7]{MSAlvim2} is sufficient. The definition of
    $\epsilon$-differential privacy in this proof is identical to the definition of
    $(\epsilon,P)$-Blowfish privacy, except for the structure of
    the adjacency relation $\sim$. As \cite[Lemma 7]{MSAlvim2} makes no assumptions about
    this relation the same arguments go through for $(\epsilon,P)$-Blowfish private channel $K$.
\end{IEEEproof}

We next transform the channel matrix such that
the diagonal corresponds to the matrix maximum
within column orbits. We construct the transformed
channel matrix $K''$ by replacing the probability
of response $j$ given $i$ given by entry $(i,j)$, with
the average of the entries in $K'$ in the orbit
of the edge $(i,j)$ under the database adjacency graph's
automorphism group $\Gamma$.

\begin{lemma} \label{lem:kprimeprime}
    Let $K'$ be an $\ell \times \ell$ channel matrix satisfying
    the conditions in Lemma \ref{lem:kprime}, and 
    $(\mathcal{X},\sim)$ be the adjacency graph. 
    Let $\Gamma$
    be a subgroup of $\Aut(\mathcal{X},\sim)$ and $K''$
    the matrix defined by:
    \begin{equation*} \label{eq:kprimeprime}
        K''_{i,j} = \frac{1}{|\Gamma|} \sum_{\sigma\in\Gamma}
            K'_{\sigma(i),\sigma(j)},\ i,j\in[\ell]\enspace.
    \end{equation*}
    Then $K''$ has the following properties:
    \begin{enumerate}[label=(\alph*)]
        \item $K''$ is a channel matrix, \ie
        $K''_{i,j}\in[0,1]$ and $\sum_{h=1}^{\ell}
            K''_{i,h}=1$ for all $i,j\in[\ell]$;
        \item Each diagonal entry of $K''$ is the maximum in its column $j\in[\ell]$:
            $K''_{j,j} = \max_{i\in[\ell]} K''_{i,j}$;
            moreover $K''_{i,i}=K''_{h,h}$ whenever $i,h$ are in the same
            $\Gamma$-orbit on $\mathcal{X}$;
            if in addition all diagonal entries of $K'$ are equal
            (and hence are maximum entries of $K$), then so too are all diagonal entries
            of $K''$;
        \item $K''$ satisfies $(\epsilon, P)$-Blowfish privacy, 
            $K''_{i,j} \leq e^\epsilon K''_{h,j}$ for all $i,j,h\in[\ell]$
            such that $i\sim h$; and
        \item $H_\infty^{K''}(X|Z) = H_\infty^{K'}(X|Z)$ if $X$ has the
            uniform distribution over $\mathcal{X}$.
    \end{enumerate}
\end{lemma}

\begin{IEEEproof}
For $i\in[\ell]$ we have,
            \begin{align*}
                \sum_{j=1}^{\ell} K''_{i,j}
		    =& \sum_{j=1}^{\ell} \frac{1}{|\Gamma|}\sum_{\sigma\in\Gamma}
                        K'_{\sigma(i),\sigma(j)} \\
		     =& \frac{1}{|\Gamma|} \sum_{\sigma\in\Gamma} \sum_{k=1}^{\ell}
                        K'_{\sigma(i),k}
		     = \frac{1}{|\Gamma|} \sum_{\sigma\in\Gamma} 1
		     = 1\enspace.
            \end{align*}
            The second equality holds since for $\sigma\in\Gamma$, we have
            $\{\sigma(1),\ldots,\sigma(\ell)\}=[\ell]$
            and the second last equality holds because $K'$ is a channel matrix
            by Lemma~\ref{lem:kprime}(a). This proves property (a).

            Let $j,h\in[\ell]$ and $\sigma\in\Gamma$. Since $K'$ satisfies
            the conditions of Lemma \ref{lem:kprime}, the maximum entry in its $\sigma(j)^{th}$
            column must be $K'_{\sigma(j),\sigma(j)}$. So,
            \begin{align*}
		    K''_{j,j} &= \frac{1}{|\Gamma|} \sum_{\sigma\in\Gamma} K'_{\sigma(j),\sigma(j)}
		    \geq \frac{1}{|\Gamma|}\sum_{\sigma\in\Gamma} K'_{\sigma(h),\sigma(j)}
		    = K''_{h, j}\enspace.
            \end{align*}
            Hence each diagonal entry of $K''$ is the maximum in its column.
            For $i\in[\ell]$ we have,
            \begin{align*}
		    K''_{i,i} =& \frac{1}{|\Gamma|}\sum_{\sigma\in\Gamma} K'_{\sigma(i),\sigma(i)} \\
		    =& \frac{1}{|\Gamma|}\sum_{k\in\Gamma(i)} \sum_{\sigma\in\Gamma_{i\rightarrow k}} K'_{k,k}
		    = \frac{1}{|\Gamma|} \sum_{k\in\Gamma(i)} |\Gamma_{i\rightarrow k}| K'_{k,k}\enspace.
            \end{align*}
            From Lemma \ref{lem:autsizes} we have $|\Gamma_{i\rightarrow k}|/|\Gamma|=1/|\Gamma(i)|$,
            hence
            \begin{equation*}
                \frac{1}{|\Gamma|} \sum_{k\in\Gamma(i)} |\Gamma_{i\rightarrow k}| K'_{k,k}
                    = \sum_{k\in\Gamma(i)} \frac{1}{|\Gamma(i)|} K'_{k,k}\enspace.
            \end{equation*}
            Therefore $K''_{i,i}=\sum_{k\in\Gamma(i)} K'_{k,k}/|\Gamma(i)|$. Hence if $i,h\in[\ell]$
            such that $\Gamma(i)=\Gamma(h)$, \ie $i$ and $h$ are in the same orbit,
            then $K''_{i,i}=K''_{h,h}$.
            Furthermore, if all diagonal entries of $K'$ are equal (and thus equal to the maximum
            element of $K'$), then so too are all diagonal entries of $K''$, for all $i\in[\ell]$:
            \begin{align*}
		     K''_{i,i} =& \sum_{k\in\Gamma(i)} \frac{1}{|\Gamma(i)|} K'_{k,k} \\
		    =& \sum_{k\in\Gamma(i)} \frac{1}{|\Gamma(i)|} \max_{h,j\in[\ell]} K'_{h,j}
		    = \max_{h,j\in[\ell]} K_{h,j}'\enspace.
            \end{align*}
        This establishes the last part of property (b).
        
        Let $i,h\in[\ell]$ be such that $i\sim h$.
            First note that for all $\sigma\in\Gamma$, $\sigma(i) \sim \sigma(h)$ from the definition
            of an automorphism. Also note that $K'$
            satisfies $(\epsilon,P)$-Blowfish privacy, so for all $j\in[\ell]$
            we have $K'_{\sigma(i),\sigma(j)} \leq e^\epsilon K'_{\sigma(h),\sigma(j)}$. So,
            \begin{align*}
                K''_{i,j}
		    &= \frac{1}{|\Gamma|}\sum_{\sigma\in\Gamma} K'_{\sigma(i),\sigma(j)}
		    \leq \frac{1}{|\Gamma|}\sum_{\sigma\in\Gamma} e^\epsilon
                        K'_{\sigma(h),\sigma(j)}
		    = e^\epsilon K''_{h,j}.
            \end{align*}
            Therefore $K''$ satisfies $(\epsilon,P)$-Blowfish privacy and property (c).
        
        We now prove the final property (d), which is rather more involved than the previous three properties of $K''$.
            Denote $r$ the number of $\Gamma$-orbits on $\mathcal{X}$. Let
            $\mathcal{X}_1,\ldots,\mathcal{X}_r$ be these orbits, with cardinalities
            $c_1,\ldots,c_r$.
            Note that $\{\mathcal{X}_1,\ldots,\mathcal{X}_r\}$ is then a partition
            of $\mathcal{X}$, and that for each $s\in[r]$ we have that $\mathcal{X}_s$
            is $\Gamma$-vertex transitive.
            Fix $i_s\in\mathcal{X}_s$, the generator of $\mathcal{X}_s$ so that
            $\Gamma(i_s)=\mathcal{X}_s$. Then $c_s=|\mathcal{X}_s|=\frac{|\Gamma|}{|\Gamma_{i_s}|}$
            by the orbit-stabiliser theorem.
            
            Denote by $[\Gamma:\Gamma_{i_s}]$ the set of left cosets of $\Gamma_{i_s}$ in $\Gamma$.
            Choose $\{\sigma_{s1},\ldots,\sigma_{sc_s}\}$ to be a set of representatives of
            $[\Gamma:\Gamma_{i_s}]$, so that $\Gamma=\cup_{t=1}^{c_s} \sigma_{st}\Gamma_{i_s}$.
            For all $t\in[c_s]$ denote $i^*_{st}=\sigma_{st}(i_s)$.
            For all $\gamma\in\Gamma_{i_s}$ and $t\in[c_s]$ we have
            $(\sigma_{st}\circ\gamma)(i_s)=\sigma_{st}(i_s)$, so
            \begin{align*}
		    \mathcal{X}_s &= \Gamma(i_s) 
		    = \bigcup_{t=1}^{c_s} \{(\sigma_{st}\circ\gamma)(i_s)
                        \:|\: \gamma\in\Gamma_{i_s}\} \\
		    &= \bigcup_{t=1}^{c_s} \{\sigma_{st}(i_s)\}  
		    = \bigcup_{t=1}^{c_s} \{i^*_{st}\} 
		    = \{i^*_{s1},\ldots,i^*_{sc_s}\}\enspace.
            \end{align*}
            Since $\mathcal{X}_s$ is $\Gamma$-vertex transitive,
            $\sum_{\sigma\in\Gamma}K_{\sigma(j),\sigma(j)}$ is independent of the choice
            of $j\in\mathcal{X}_s$. Fixing $k\in\mathcal{X}_s$, for all $j\in\mathcal{X}_s$
            we have,
            \begin{align*}
                \sum_{j\in\mathcal{X}_s} \sum_{\sigma\in\Gamma} K'_{\sigma(j),\sigma(j)}
		    =& c_s \sum_{\sigma\in\Gamma} K'_{\sigma(k),\sigma(k)} \\
		    =& c_s \sum_{t=1}^{c_s} \sum_{\sigma\in\Gamma_{k\rightarrow i^*_{st}}}
                        K'_{i^*_{st},i^*_{st}}\enspace.
            \end{align*}
            From Lemma \ref{lem:autsizes} we know that
            $|\Gamma_{k\rightarrow i^*_{st}}|=|\Gamma_{i^*_{st}}|$, and so,
            \begin{align*}
                    c_s \sum_{t=1}^{c_s} \sum_{\sigma\in\Gamma_{k\rightarrow i^*_{st}}}
                        K'_{i^*_{st},i^*_{st}}
		    =& c_s |\Gamma_{k\rightarrow i^*_{st}}| \sum_{t=1}^{c_s} 
                        K'_{i^*_{st},i^*_{st}} \\
		    =& c_s |\Gamma_{i^*_{st}}| \sum_{t=1}^{c_s}
                        K'_{i^*_{st},i^*_{st}}\enspace.
            \end{align*}
            Since $i^*_{st} = \sigma_{st}(i_s)$, we know that
            $\Gamma_{i^*_st\rightarrow i_s}\neq \emptyset$. Thus Lemma \ref{lem:stabsize}
            yields $|\Gamma_{i^*_{st}}|=|\Gamma_{i_s}|$. Also recall that
            $c_s=|\Gamma|/|\Gamma_{i_s}|$ and $\mathcal{X}_s=\{i^*_{s1},\ldots,i^*_{sc_s}\}$.
            Therefore,
            \begin{align*}
                c_s |\Gamma_{i^*_{st}}| \sum_{t=1}^{c_s}
                    K'_{i^*_{st},i^*_{st}}
		    =& c_s |\Gamma_{i_s}| \sum_{t=1}^{c_s} 
                    K'_{i^*_{st},i^*_{st}}\\
		 =& |\Gamma| \sum_{t=1}^{c_s}
                    K'_{i^*_{st},i^*_{st}} 
		    = |\Gamma|\sum_{j\in\mathcal{X}_s} K'_{j,j}\enspace.
            \end{align*}
            So $\sum_{j\in\mathcal{X}_s} \sum_{\sigma\in\Gamma} K'_{\sigma(j),\sigma(j)}
            = |\Gamma|\sum_{j\in\mathcal{X}_s} K'_{j,j}$.
            Now consider,
            \begin{align*}
                \sum_{j=1}^{\ell} K''_{j,j}
		    &= \sum_{s=1}^{r} \sum_{j\in\mathcal{X}_s} K''_{j,j} \\
		    &= \sum_{s=1}^{r} \sum_{j\in\mathcal{X}_s} \left(
                        \frac{1}{|\Gamma|}\sum_{\sigma\in\Gamma} K'_{\sigma(j),\sigma(j)}
                        \right) \\
		    &= \sum_{s=1}^{r} \sum_{j\in\mathcal{X}_s} K'_{j,j} \\
		    &= \sum_{j=1}^{\ell} K'_{j,j}\enspace.
            \end{align*}
            So the sum of the diagonals of $K''$ is equal to the sum of diagonals of $K'$.
            We know from Lemma \ref{lem:Hofmatrix} that $H_\infty^K(X|Z)$ is a function
            of the sum of the maximum entries in each column of $K$. In addition, we know
            that both $K'$ and $K''$ attain a maximum for each column in the diagonal
            from Lemmas~\ref{lem:kprime}(b) and~\ref{lem:kprimeprime}(b). Therefore we have both
            \begin{align*}
                H_\infty^{K''}(X|Z) &= -\log \frac{1}{\ell} \sum_{j=1}^{\ell}
                        \max_{i} K''_{i,j} 
                    = -\log \frac{1}{\ell} \sum_{j=1}^{\ell} K''_{j,j} \\
                H_\infty^{K'}(X|Z) &= -\log \frac{1}{\ell} \sum_{j=1}^{\ell}
                        \max_{i} K'_{i,j} 
                    = -\log \frac{1}{\ell} \sum_{j=1}^{\ell} K'_{j,j}\enspace.
            \end{align*}
            We have shown that $\sum_{j=1}^{\ell} K''_{j,j} = \sum_{j=1}^{\ell} K'_{j,j}$,
            and as such $H_\infty^{K''}(X|Z)=H_\infty^{K'}(X|Z)$. So property (d) is satisfied
            by $K''$.
\end{IEEEproof}

\subsection{Proof of Theorem~\ref{thm:minent}}

    Let $\epsilon, P$ satisfy the theorem's conditions, 
    $(\mathcal{X},\mathcal{Z},K)$
    be an $(\epsilon,P)$-Blowfish-private channel. Assume $X$ is uniformly distributed over $\mathcal{X}$.
    From Lemmas~\ref{lem:kprime} and~\ref{lem:kprimeprime} we know that we can transform
    the $\ell\times p$ channel matrix $K$ into an $\ell \times \ell$ channel matrix $K''$ satisfying
    Lemma \ref{lem:kprimeprime}'s conditions.
    
    Let $q$ be the number of connected components of  $(\mathcal{X},\sim)$. These components,
    $\{\mathcal{X}^{(1)},\ldots,\mathcal{X}^{(q)}\}$, partition $\mathcal{X}$.
    From Lemmas~\ref{lem:Hofmatrix} and~\ref{lem:kprimeprime}(b,d) we know that,
    \begin{align*}
	    H_\infty^{K}(X|Z) =& H_\infty^{K''}(X|Z) \\
	    =& -\log \frac{1}{\ell} \sum_{j=1}^{\ell} \max_i K''_{i,j}
	    = -\log \frac{1}{\ell} \sum_{j=1}^{\ell} K''_{j,j}\enspace.
    \end{align*}
    Let $t\in[q]$, and let $i,j,k\in\mathcal{X}^{(t)}$. For elements $i$ and $h$ in a connected component,
    the Blowfish privacy definition for a channel matrix (\ref{eq:channelblowfish}) can be
    extended to $K''_{h,j} \leq e^{\epsilon d(i,h)} K''_{i,j}$.
    In particular, letting $h=j$, this yields $K''_{j,j} \leq e^{\epsilon d(i,j)} K''_{i,j}$. 
    For each $t\in[q]$ select $i_t\in\mathcal{X}^{(t)}$. Also denote the {\it diameter} of the connected component $\mathcal{X}^{(t)}$ by
    $d_t=\operatorname{Diam}(\mathcal{X}^{(t)}) := \max_{i,j\in\mathcal{X}^{(t)}} d(i,j)$. Now,
    \begin{align*}
        \sum_{j=1}^\ell K''_{j,j}
	    &= \sum_{t=1}^q \sum_{j\in\mathcal{X}^{(t)}} K''_{j,j} \\
	    &\leq \sum_{t=1}^q \sum_{j\in\mathcal{X}^{(t)}} e^{\epsilon d(i_t,j)} K''_{i_t,j} \\
	    &\leq \sum_{t=1}^q e^{\epsilon d_t} \sum_{j\in\mathcal{X}^{(t)}} K''_{i_t,j}\enspace.
    \end{align*}
    Since $K''$ is a channel matrix $K''_{i,j}\geq 0$ for all $i,j\in[\ell]$, so
    $\sum_{j\in\mathcal{X}^{(t)}} K''_{i_t,j} \leq \sum_{j=1}^\ell K''_{i_t,j}$.
    Also since $K''$ is a channel matrix, $\sum_{j=1}^\ell K''_{i_t,j} = 1$. So,
    \begin{equation*}
        \sum_{t=1}^q e^{\epsilon d_t} \sum_{j\in\mathcal{X}^{(t)}} K''_{i_t,j}
            \;\leq\; \sum_{t=1}^q e^{\epsilon d_t} \sum_{j=1}^\ell K''_{i_t,j}
            \;=\; \sum_{t=1}^q e^{\epsilon d_t}\enspace.
    \end{equation*}
    Then we have $\sum_{j=1}^\ell K''_{j,j} \leq \sum_{t=1}^q e^{\epsilon d_t}$, hence
    \begin{equation*}
        H_\infty^K(X|Z) \;=\; -\log \frac{1}{\ell} \sum_{j=1}^{\ell} K''_{j,j}
            \;\geq\; -\log \frac{1}{\ell} \sum_{t=1}^q e^{\epsilon d_t}\enspace.
    \end{equation*}
    
\subsection{Proof of Main Theorem~\ref{thm:infleak}}
    Assume $\epsilon$ and $P$ satisfy the conditions in the theorem statement. Let
    $(\mathcal{X}, \mathcal{Z}, K)$ be an $(\epsilon,P)$-Blowfish-private channel.
    Note that $X$ can have any prior over $\mathcal{X}$ and is not required to be uniformly distributed.
    Assume $X^\text{uniform}$ has the uniform distribution over $\mathcal{X}$.
    From Theorem~\ref{thm:minent} we know that $H_\infty^{K}(X^\text{uniform}|Z)
    \geq -\log \frac{1}{\ell} \sum_{t=1}^q e^{\epsilon d_t}$,
    with $q$ the number
    of connected components of $\mathcal{X}$, $d_t$ the diameter of component $\mathcal{X}^{(t)}$. Observe that min-entropy of $X^\text{uniform}$ is 
     $H_\infty^{K}(X^\text{uniform}) = -\log \max_{x\in\mathcal{X}} p(x) = -\log 1/\ell
    = \log \ell$. So,
    \begin{align*}
        I_\infty^K (X^\text{uniform};Z)
	    &= H_\infty^{K}(X^\text{uniform}) - H_\infty^{K}(X^\text{uniform}|Z) \\
	    &\leq \log \ell + \log \frac{1}{\ell} \sum_{t=1}^q e^{\epsilon d_t}
	    = \log \sum_{t=1}^q e^{\epsilon d_t}.
    \end{align*}
    Braun \etal~\cite[Proposition 5.1]{braun2009leakage} demonstrate that maximum leakage is attained over
    the uniform distribution, \ie $I^K_\infty(X;Z)\leq I^K_\infty(X^\text{uniform};Z)$. Therefore
    \begin{equation*}
        I_\infty^K (X;Z)
            \;\leq\; I_\infty^K (X^\text{uniform};Z)
            \;\leq\; \log \sum_{t=1}^q e^{\epsilon d_t}\enspace.
    \end{equation*}

\subsection{Relationship Between the Graph Automorphism Groups}
Our proof uses the (full) automorphism group
for the database adjacency graph, \ie $\Aut(\mathcal{I},\sim)$.
As discussed in Section~\ref{sec:bg-blowfish}, the adjacency graph $(\mathcal{I},\sim)$ is defined in terms of the secret graph
$(\mathcal{T},E)$;
$D\sim D'$ if $D$ and $D'$ are \textit{minimally secretly different}
(Definition~\ref{def:minsecretelydiff}).

To expand on the relationship between these graphs we demonstrate the
relationship between their automorphism groups.

\begin{theorem} \label{thm:secretaut}
If all databases are permissible (\ie $\mathcal{I}=\mathcal{T}^n$)
and $\varphi_1,\ldots,\varphi_n\in\Aut(\mathcal{T},E)$ are
automorphisms for the secret graph then
\begin{gather*}
    \sigma(D) = \sigma((t_1,\ldots,t_n))
    := \left(\varphi_1(t_1),\ldots,\varphi_n(t_n)\right)
\end{gather*}
is an automorphism for the database adjacency graph.
\end{theorem}

\begin{IEEEproof}
Recall that we have
$\tdiff(D,D')=\left\{
        (i,u,v)\in[n]\times\mathcal{T}^2    \mid
        u=D_i,v=D'_i,u\neq v
    \right\}$.
Since each $\varphi_i$ is a permutation on $\mathcal{T}$
for $u,v\in\mathcal{T}$ we have $u=v$ if and only
if $\varphi_i(u)=\varphi_i(v)$. Hence, $\tdiff(D,D')$
and $\tdiff(\sigma(D),\sigma(D'))$ are in bijection.
Similarly, the secret difference is defined as
$\sdiff(D,D') = \left\{ (i, u, v)\in \tdiff(D,D') \mid
        (u,v)\in E
    \right\}$.
If $(i,u,v)\in\sdiff(D,D')$ then $D_i = u$, $D'_i=v$, and $(u,v)\in E$.
Since $\varphi_i$ a graph automorphism, $(\varphi_i(u),\varphi_i(v))\in E$.
So $(i,\varphi_i(u),\varphi_i(v))\in \sdiff(\sigma(D),\sigma(D'))$.
Since the $\varphi_i$s are bijective, applying this argument in reverse
demonstrates that $\sdiff(D,D')$ and $\sdiff(\sigma(D),\sigma(D'))$
are in bijection.

To show that $\sigma$ defines an automorphism for $(\mathcal{I},\sim)$
we need to show that $\sigma$ is a permutation on $\mathcal{I}$
(\ie a bijection from $\mathcal{I}$ to $\mathcal{I}$)
such that $D\sim D'$ implies $\sigma(D)\sim\sigma(D')$. The
first point follows from the definition of $\sigma$ as
the composition of element-wise permutations. The second
point follows from the definition
of $\sim$ (Definition~\ref{def:minsecretelydiff}),
using the fact that $\tdiff(D,D')$
and $\tdiff(\sigma(D),\sigma(D'))$
are in bijection, as are $\sdiff(D,D')$ and $\sdiff(\sigma(D),\sigma(D'))$.
\end{IEEEproof}

\begin{remark}\label{rem:secretaut}
The graph automorphisms from Theorem~\ref{thm:secretaut}
don't generate all of $\Aut(\mathcal{I},\sim)$.
Consider again $\mathcal{I}=\mathcal{T}^n$ and
functions $\sigma:\mathcal{I}\rightarrow\mathcal{I}$
which are permutations of the elements in $D$.
That is, suppose $\pi:[n]\rightarrow[n]$ is a permutation,
then
\begin{gather*}
    \sigma((t_1,\ldots,t_n)) = (t_{\pi(1)},\ldots,t_{\pi(n)})
\end{gather*}
defines an automorphism for $\Aut(\mathcal{I},\sim)$. This
automorphism is not composed of element-wise transformations
and thus cannot be generated by Theorem~\ref{thm:secretaut}.
\end{remark}

\begin{example}\label{eg:cyclecomplete-aut}
    Consider the secret graphs from Example~\ref{eg:cyclecomplete},
    which are (a) the cyclic graph on $m$ vertices, $C_m$, and (b)
    the complete graph on $m$ vertices, $K_m$.
    The automorphism group $\Aut(C_m)$ is isomorphic
    to the dihedral group $D_m$ of order $2m$, containing
    rotational and reflective symmetries of a regular $m$-gon.
    The automorphism group $\Aut(K_m)$ is the symmetric group
    $\operatorname{Sym}(m)$ of all bijective functions from
    $[m]$ to $[m]$.
\end{example}

\section{Conclusions}

This paper considers leading frameworks within two parallel threads
of research on privacy-preserving aggregate or model release: min-entropy leakage within quantitative
information flow and Blowfish privacy within differential privacy. The only known link
between threads is a bound
of Alvim \etal~\cite{MSAlvim2} on min-entropy leakage by differential privacy which requires strong
symmetry assumptions of the database adjacency graph, of distance regularity and vertex transitivity. Adjacency
graphs under Blowfish privacy are arbitrary by design: the appealing property of
Blowfish is its relaxation of adjacency for capturing public knowledge of underlying
data. It is therefore
interesting to understand how the graph structure of Blowfish semantic privacy
policies---represented by adjacency graphs---bounds min-entropy leakage. We overcome this
challenge by organising analysis around vertex-transitive automorphism
orbits. 
Our results relate these two important frameworks, and shed light on the structure
of Blowfish privacy policies and their implications. 

Noting that differential privacy and min-entropy leakage are well defined over continuous sets of tuples, while our results assume finite $\mathcal{T}$, it is an interesting open question as to whether our symmetrisation argument extends relations on uncountable $\mathcal{T}$.

\section*{Acknowledgements}

We are grateful to the reviewers for suggesting the examination of cyclic secret graphs (the subject of Examples~\ref{eg:cyclecomplete} and~\ref{eg:cyclecomplete-aut}), relating the automorphism groups of the secret and database adjacency graphs (the subject of Theorem~\ref{thm:secretaut}, Remark~\ref{rem:secretaut} and Example~\ref{eg:cyclecomplete-aut}), and for offering many more helpful comments that improved this work.

\ifCLASSOPTIONcaptionsoff
  \newpage
\fi

\bibliographystyle{IEEEtran}
\bibliography{IEEEabrv,references}  %

\newpage

\begin{IEEEbiographynophoto}{Tobias Edwards}
    received the B.S. degree in computing and software systems and the Diploma in pure mathematics from The University of Melbourne in 2018 followed by the Graduate Diploma (Advanced) in statistics and stochastic processes in 2021, also from The University of Melbourne. During his undergraduate studies Tobias spent time as a visiting student at Carnegie Mellon University, focussing on algorithmic analysis and machine learning. He has recently worked in industrial machine learning with Xero and is currently working on global equity market investments and technology with Catalyst Funds Management.
\end{IEEEbiographynophoto}

\begin{IEEEbiographynophoto}{Benjamin I. P. Rubinstein}
    received the PhD (EECS) from UC Berkeley in 2010. He 
    has worked at Microsoft Research Silicon Valley, IBM Research Australia, and since 2013 he
    has been with the School of Computing and Information Systems at The University of
    Melbourne where he is currently a Professor. His research interests span machine learning, security \& privacy, and databases---in particular adversarial machine learning, differential privacy, and record linkage. He has completed contract research with Facebook, the Australian Bureau of Statistics, National Australia Bank, DSTG, CSIRO Data61, Transport for NSW, and OVIC in these areas. From 2016 to 2018 he held a DECRA from the Australian Research Council. Since 2021 he is co-lead of the joint MURI-AUSMURI Cybersecurity Assurance for Teams of Computers and Humans (CATCH) funded by U.S. ARO and the Australian Department of Industry, Science, Energy and Resources. 
\end{IEEEbiographynophoto}

\begin{IEEEbiographynophoto}{Zuhe Zhang}
    received the PhD degree in mathematics from The University of Melbourne in 2017. His doctoral research focused on Bayesian inference under differential privacy and the application of combinatorics in statistical mechanics. He is currently with the Australian Taxation Office, working on understanding documents using transformer-based models.
\end{IEEEbiographynophoto}

\begin{IEEEbiographynophoto}{Sanming Zhou}
    received PhD with Distinction from the University of Western Australia in 2000. Since then he has been working in the School of Mathematics and Statistics, The University of Melbourne, where he is currently a Professor. His research interest spans from pure to applied aspects of discrete mathematics, including algebraic graph theory, combinatorial optimization, random graph processes and various problems from theoretical computer science. He has published over 120 research papers mostly in high quality international journals in these areas. In 2003 he was awarded a Kirkman Medal by the Institute of Combinatorics and its Applications for his outstanding achievements in research. From 2012 to 2015 he held a Future Fellowship of the Australian Research Council. Since 2018 he has been serving as Co-Editor-in-Chief of the Australasian Journal of Combinatorics.   
\end{IEEEbiographynophoto}

\vfill

\end{document}

%% file: figures/secrets.tikz
\begin{tikzpicture}[node distance = 1.5cm]

    \node [draw, circle] (1) {1};
    \node [draw, circle, right of=1] (2) {2};
    \node [draw, circle, right of=2] (3) {3};
    \node [draw, circle, right of=3] (4) {4};
    \path [draw] (1) -- (2);
    \path [draw] (2) -- (3);
    \path [draw] (3) -- (4);
    \path (1) edge[bend right, densely dotted] node [left] {} (3);
    \path (2) edge[bend right, densely dotted] node [left] {} (4);
    \path (1) edge[bend left, loosely dashed] node [left] {} (4);
    
\end{tikzpicture}

%% file: figures/adjacency.tikz
\begin{tikzpicture}[node distance = 2cm]

    \node [draw, circle] (11) {1,1};
    \node [draw, circle, right of=11] (12) {1,2};
    \node [draw, circle, right of=12] (13) {1,3};
    \node [draw, circle, right of=13] (14) {1,4};

    \node [draw, circle, below of=11] (21) {2,1};
    \node [draw, circle, right of=21] (22) {2,2};
    \node [draw, circle, right of=22] (23) {2,3};
    \node [draw, circle, right of=23] (24) {2,4};

    \node [draw, circle, below of=21] (31) {3,1};
    \node [draw, circle, right of=31] (32) {3,2};
    \node [draw, circle, right of=32] (33) {3,3};
    \node [draw, circle, right of=33] (34) {3,4};

    \node [draw, circle, below of=31] (41) {4,1};
    \node [draw, circle, right of=41] (42) {4,2};
    \node [draw, circle, right of=42] (43) {4,3};
    \node [draw, circle, right of=43] (44) {4,4};

    \path [draw] (11) -- (12);
    \path [draw] (12) -- (13);
    \path [draw] (13) -- (14);

    \path [draw] (21) -- (22);
    \path [draw] (22) -- (23);
    \path [draw] (23) -- (24);

    \path [draw] (31) -- (32);
    \path [draw] (32) -- (33);
    \path [draw] (33) -- (34);

    \path [draw] (41) -- (42);
    \path [draw] (42) -- (43);
    \path [draw] (43) -- (44);

    \path [draw] (11) -- (21);
    \path [draw] (21) -- (31);
    \path [draw] (31) -- (41);

    \path [draw] (12) -- (22);
    \path [draw] (22) -- (32);
    \path [draw] (32) -- (42);

    \path [draw] (13) -- (23);
    \path [draw] (23) -- (33);
    \path [draw] (33) -- (43);

    \path [draw] (14) -- (24);
    \path [draw] (24) -- (34);
    \path [draw] (34) -- (44);

    \path (11) edge[bend right=45, densely dotted] node [left] {} (31);
    \path (21) edge[bend right=45, densely dotted] node [left] {} (41);

    \path (12) edge[bend right=45, densely dotted] node [left] {} (32);
    \path (22) edge[bend right=45, densely dotted] node [left] {} (42);

    \path (13) edge[bend right=45, densely dotted] node [left] {} (33);
    \path (23) edge[bend right=45, densely dotted] node [left] {} (43);

    \path (14) edge[bend right=45, densely dotted] node [left] {} (34);
    \path (24) edge[bend right=45, densely dotted] node [left] {} (44);

    \path (11) edge[bend right=45, densely dotted] node [left] {} (13);
    \path (12) edge[bend right=45, densely dotted] node [left] {} (14);

    \path (21) edge[bend right=45, densely dotted] node [left] {} (23);
    \path (22) edge[bend right=45, densely dotted] node [left] {} (24);

    \path (31) edge[bend right=45, densely dotted] node [left] {} (33);
    \path (32) edge[bend right=45, densely dotted] node [left] {} (34);

    \path (41) edge[bend right=45, densely dotted] node [left] {} (43);
    \path (42) edge[bend right=45, densely dotted] node [left] {} (44);

    \path (11) edge[bend left=25, loosely dashed] node [left] {} (41);
    \path (12) edge[bend left=25, loosely dashed] node [left] {} (42);
    \path (13) edge[bend left=25, loosely dashed] node [left] {} (43);
    \path (14) edge[bend left=25, loosely dashed] node [left] {} (44);

    \path (11) edge[bend left=25, loosely dashed] node [left] {} (14);
    \path (21) edge[bend left=25, loosely dashed] node [left] {} (24);
    \path (31) edge[bend left=25, loosely dashed] node [left] {} (34);
    \path (41) edge[bend left=25, loosely dashed] node [left] {} (44);

  \end{tikzpicture}

%% file: figures/leakageplot.tikz
\begin{tikzpicture}
    \begin{axis}[
        ymin=0,
        xmin=0,
        xmax=20,
        ylabel={$I_\infty$ Upper Bound},
        xlabel=$n$,
        xtick pos=left,
        ytick pos=left,
        yticklabels={0, 0, $20\epsilon$, $40\epsilon$, $60\epsilon$}
    ]

    \addplot[domain=0:20]{x};
    \addplot[domain=0:20]{2*x};
    \addplot[domain=0:20]{3*x};
    
    \node at (150, 100) {$\theta=3$};
    \node at (150, 250) {$\theta=2$};
    \node at (150, 380) {$\theta=1$};

    \end{axis}
\end{tikzpicture}

%% file: figures/cyclesecrets.tikz
\begin{tikzpicture}[node distance = 1.5cm]

    \node[draw=none,minimum size=4cm,regular polygon,regular polygon sides=6] (ngon) {};
    
    \draw [domain=-120:180] plot ({2*cos(\x)}, {2*sin(\x)});
    \draw [domain=180:240, dashed] plot ({2*cos(\x)}, {2*sin(\x)});
    
    \node [draw, circle, fill=white, minimum size=0.5cm, label=above right:{1}]
        at (ngon.corner 1) {};
    \node [draw, circle, fill=white, minimum size=0.5cm, label=above left:{2}]
        at (ngon.corner 2) {};
    \node [draw, circle, fill=white, minimum size=0.5cm, label=left:{3}]
        at (ngon.corner 3) {};
    \node [draw, circle, fill=white, minimum size=0.5cm, label=below left:{$m-2$}]
        at (ngon.corner 4) {};
    \node [draw, circle, fill=white, minimum size=0.5cm, label=below right:{$m-1$}]
        at (ngon.corner 5) {};
    \node [draw, circle, fill=white, minimum size=0.5cm, label=right:{$m$}]
        at (ngon.corner 6) {};
    
\end{tikzpicture}

%% file: figures/completesecrets.tikz
\begin{tikzpicture}[node distance = 1.5cm]

    \node[draw=none,minimum size=4cm,regular polygon,regular polygon sides=6] (ngon) {};
    
    \draw [domain=-120:180] plot ({2*cos(\x)}, {2*sin(\x)});
    \draw [domain=180:240, dashed] plot ({2*cos(\x)}, {2*sin(\x)});
    
    \path [draw] (ngon.corner 1) -- (ngon.corner 3);
    \path [draw] (ngon.corner 1) -- (ngon.corner 4);
    \path [draw] (ngon.corner 1) -- (ngon.corner 5);
    \path [draw] (ngon.corner 2) -- (ngon.corner 4);
    \path [draw] (ngon.corner 2) -- (ngon.corner 5);
    \path [draw] (ngon.corner 2) -- (ngon.corner 6);
    \path [draw] (ngon.corner 3) -- (ngon.corner 5);
    \path [draw] (ngon.corner 3) -- (ngon.corner 6);
    \path [draw] (ngon.corner 4) -- (ngon.corner 6);
    
    \node [draw, circle, fill=white, minimum size=0.5cm, label=above right:{1}]
        at (ngon.corner 1) {};
    \node [draw, circle, fill=white, minimum size=0.5cm, label=above left:{2}]
        at (ngon.corner 2) {};
    \node [draw, circle, fill=white, minimum size=0.5cm, label=left:{3}]
        at (ngon.corner 3) {};
    \node [draw, circle, fill=white, minimum size=0.5cm, label=below left:{$m-2$}]
        at (ngon.corner 4) {};
    \node [draw, circle, fill=white, minimum size=0.5cm, label=below right:{$m-1$}]
        at (ngon.corner 5) {};
    \node [draw, circle, fill=white, minimum size=0.5cm, label=right:{$m$}]
        at (ngon.corner 6) {};
    
\end{tikzpicture}

%% file: figures/sharpness.tikz
\begin{tikzpicture}[node distance = 2.0cm, minimum width=1.2cm, inner sep=0pt]

\node [draw, circle] (1) {$x_1$};
\node [draw, circle, below right of=1] (2) {$x_2$};
\node [draw, circle, above right of=1] (3) {$x_3$};
\node [draw, circle, below right of=3] (4) {$x_4$};

\node [draw, circle, above right of=4] (5) {$x_5$};
\node [draw, circle, right of=5] (6) {$x_6$};

\node [draw=none, right of=4] (7) {\hspace{2em} $\cdots$};

\node [draw, circle, below right of=4] (8) {$x_{2n+1}$};
\node [draw, circle, right of=8] (9) {$x_{2n+2}$};

\path [draw] (1) -- (2);
\path [draw] (1) -- (3);
\path [draw] (1) -- (4);
\path [draw] (2) -- (3);
\path [draw] (2) -- (4);
\path [draw] (3) -- (4);

\path [draw] (5) -- (6);

\path [draw] (8) -- (9);

  \end{tikzpicture}